\documentclass[conference]{IEEEtran}
\usepackage{cite}

\ifCLASSINFOpdf
   \usepackage[pdftex]{graphicx}
\else
  \usepackage[dvips]{graphicx}
\fi
\usepackage[cmex10]{amsmath}
\begin{document}
%
\title{Optimization of Bloom Filter Parameters for Practical Bloom Filter Based Epidemic Forwarding in DTNs}
\author{\IEEEauthorblockN{Ali Marandi}
\IEEEauthorblockA{Department of Computer Engineering\\Khorasgan (Isfahan) Branch\\
 Islamic Azad University\\
 Isfahan, Iran\\
Email: ali.marandi@khuisf.ac.ir}
\and
\IEEEauthorblockN{Mahdi Faghih Imani}
\IEEEauthorblockA{Department of Computer Engineering\\Science and Research Branch\\
Islamic Azad University\\
Tehran, Iran\\
Email: m.imani@srbiau.ac.ir}
\and
\IEEEauthorblockN{Kave Salamatian\\ }
\IEEEauthorblockA{Listic-PolyTech Universite de Savoie\\
France\\
Email: kave.salamatian@univ-savoei.fr
}}

%
\maketitle
\begin{abstract}
Epidemic forwarding has been proposed as a forwarding technique to achieve opportunistic communication in Delay Tolerant Networks. Even if this technique is well known and widely referred, one has to first deal with several practical problems before using it. In particular, in order to manage the redundancy and to avoid useless transmissions, it has been proposed to ask nodes to exchange information about the buffer content prior to sending information. While Bloom filter has been proposed to transport the buffer content information, up to our knowledge no real evaluation has been provided to study the tradeoff that exists in practice. In this paper we describe an implementation of an epidemic forwarding scheme using Bloom filters. Then we propose some strategies for Bloom filter management based on windowing and describe implementation tradeoffs. By simulating our proposed strategies in ns-3 both with random waypoint mobility and realistic mobility traces coming from San Francisco taxicabs, we show that our proposed strategies alleviate the challenge of using epidemic forwarding in DTNs.
\end{abstract}



%
\IEEEpeerreviewmaketitle
\section{Introduction} \label{S:intro}
Delay tolerant networks (DTNs) are accompanied with several major challenges such as intermittent and transient connectivity, volatile links and long delays. Wireless sensor networks, vehicular networks (VANETs), and spontaneous networks are examples of these networks. In most of DTN applications, no information about future contacts is available and data forwarding follows  opportunistic approaches based on store-carry and forward scheme, {\em i.e.} relay nodes store packets and carry them until an appropriate forwarding opportunity. Among all techniques proposed for DTN routing, a lot of them are not applicable when no information on future contacts is available. In these situations only epidemic forwarding is usable. Epidemic Forwarding simply consists of letting encountering nodes to exchange the information stored in their buffers. However in order to control the redundancy and to avoid useless transmissions Vahdat and Becker \cite{vahdat}  proposed to exchange between nodes a summary bitmap indicating which packets are already received in nodes. However, this idea is not practical because the nodes need to be informed of an ordered list of all messages circulating in the network in order to interpret the bitmaps. Building and diffusing this list seems impracticable specially in an asynchronous multi-source/multi-destination scenario. Therefore the proposed summary bitmaps is in fact a list of received packet IDs and its exchange can impose a relatively large overhead. Vahdat and Becker \cite{vahdat}  suggested therefore to use a {\em Bloom filter} in order to substantially reduce the space overhead associated with the summary vector. Despite this approach being known from a long time, there is a relatively small number of work that described practically how to implement Bloom filter based epidemic forwarding and discussed the involved trade-offs. The aim of this paper is to discuss this issue. 

The major challenges that one has to address in order to use epidemic forwarding in practical settings are the following. First one needs to deal with practical issues as how to define in a distributed way Bloom filters and how to achieve good transmission overheads vs. transmitted redundancy trade-offs. Moreover, one also has to deal with indicating packets received at destinations in order to not forward them and to free the space occupied by them.


In this paper we describe our implementation of Bloom filter based epidemic forwarding that is based on \emph{windowing}.
%
In order to evaluate the proposed mechanism, we run simulations with \emph{ns-3} and provide the partial results. The rest of this paper is organized as follows. We first describe Bloom filter based epidemic forwarding in Sec.~\ref{S:epidemic}. Then Sec.~\ref{S:evaluation} gives the performance evaluation of the proposed schemes. After that, Sec.~\ref{S:rw} goes over some related works and finally Sec.~\ref{S:conclusion} concludes the paper.  

\section{ Bloom Filter based Epidemic Forwarding}  \label{S:epidemic}
As explained in the introduction, even if the idea of using a Bloom filter to reduce the overhead of exchanging the buffer contents upon nodes encounter, goes back to the initial paper on epidemic forwarding \cite{vahdat} and it has been reused frequently in the literature, however, almost nobody explained how to translate this concept into a practical scheme and what are the issues and challenges in this way. In this section, we will describe the existing issues and we will propose solutions to solve them. 
\subsection{Issues and challenges}
DTNs are networks that are characterized by the fact that node encounter happens randomly so that finding a deterministic path from the source of information to its destination is not feasible. Whenever one of two encountering nodes is the destination of a packet sitting in the buffer of the other node, a final delivery occurs. When no one of the nodes are the final destination of an exchanged packet, the receiving node acts as a relay. Because of the disconnectivity, nodes in DTNs have to transport several packets in the buffer in order to deliver them later.  

Basically DTN routing consists of deciding based on the characteristic of an encountered node which packets to forward to it. In classical routing, the decision about the next node to forward a received packet is made based on the destination of the packet, {\em i.e.} a packet is forwarded to a node that is defined by the routing table to be in the path for reaching the final destination. However in DTNs the situation is more complex. When the nodes mobility is unpredictable, {\em i.e.}, one cannot predict upcoming contacts of an encountered node, the identity of an encountered node is not useful to decide if it is more likely to help in forwarding a message. In such situation Epidemic Forwarding is used. In Epidemic Forwarding whenever two nodes encounter they make the relaying decision without considering the packets destination (besides when the encountered node is the final destination of a packet) but only based on their buffer content. In particular, when two nodes become in contact they should avoid sending packets that another node has already received. 

The above description defines four main challenges and issues for Epidemic forwarding in DTNs:
\begin{enumerate}
	\item {\bf Node buffer comparison}: in order to avoid exchanging redundant packets (packets that the receiving node has already received) in epidemic forwarding, the two encountering nodes have to inform each other of the content of their buffer. In their seminal paper, Vahdat and Becker \cite{vahdat}  proposed to use a summary bitmap indicating which packets are already received in nodes and to exchange this summary between nodes. However, this idea is not applicable in practice as the nodes need to have access to an ordered list containing all messages circulating in the network in order to build and interpret the bitmaps. 
	Exchanging this list can impose a relatively large overhead especially when the contact time between nodes is short. For this reason \cite{vahdat}  suggested to use a {\em Bloom filter} in order to substantially reduce the space overhead associated with the summary vector.
	\item {\bf Buffer management}: as explained before in DTN scenarios, nodes have to store a large number of packets in their buffer. However the buffer space is limited and a node will have to decide which packet to store in its buffer. This question is essential for congestion control in DTNs that falls outside the scope of this paper that mainly targets the issues relative to the use of ``Bloom filter" in DTNs. Nonetheless, Bloom filter has important impacts on Buffer management and \emph{vice-versa}. We will develop later on this last point. 
	\item {\bf Reception Acknowledgment}: Buffer space and connectivity bandwidth, \emph{i.e.}, the available capacity of transmission during a contact between nodes, are the major resources that have to be used efficiently in DTNs. A trivial way to improve the utilization of this two important resources is to inform nodes in the network about the packets that have reached their destinations. This information enables nodes to free the space allocated to these packets and ensure that they do not waste the scarce connectivity bandwidth by being reforwarded. For achieving this purpose, we need to implement a feedback scheme that will signal the reception of packets. 
	\item{\bf Neighborhood management}: because of mobility the neighborhood of each node becomes very dynamic. However, the information forwarding strongly depends on the neighborhood of the node and the knowledge that each node has about the content in the buffer of its neighbors. Neighborhood management has therefore strong impact on the performance of DTN schemes. While this is a trivial fact, however this last issue is rarely addressed in the literature. In particular a fundamental question is how to decide if two nodes are in reach of each other. In this paper we will assume that there is no particular support from lower layers (in form of synchronization or loss of synchronization) and neighborhood detection and management is only done through message exchange.
\end{enumerate}
All the above four issues will be addressed in this paper. In the next sections we will describe the system architecture, the Bloom filter management and in particular 3 strategies that will address different issues. 

\subsection{System architecture}  \label{S:architecture}
Let us assume that each packet injected in the network is identified by a combination of a uniquely assigned source node ID defined over 16 bits, and a serial index assigned locally by the source node over 16 bits. Meaning that each packet is identified by a 32 bits packet ID that uniquely identifies each packet in the DTN and can be generated asynchronously in each node.

We assume that each node in the DTN maintains a buffer with a capacity of $N$ packets containing packets that the node will forward. In addition to this buffer several lists are maintained.  A \emph{neighbors} list that contains the information related to nodes that are known to be in the neighborhood of the node. The \emph{neighbors} list contains for each neighbor the last Bloom filters provided by the node neighbor. A \emph{destReceived} list that contains the set of packets that are known to have been received at destinations. This last list is used for signaling the reception of packets at their destination and enables nodes to remove from their buffer packets received at destination.
\subsection{Neighborhood management}
As explained before neighborhood management is a fundamental issue of DTNs. We are not assuming in this paper any connectivity indication from lower layers, and all neighborhood management should be done through message exchanges. We therefore assume that whenever node A receives a message from node B, it is in its neighborhood and can receive messages sent from A. A node is removed from neighborhood if no packet is received from it during a disconnection timer duration. We moreover assume that nodes send periodically (with a period less than the disconnection timer) a beacon message to keep connections and neighborhood alive. 

The above scheme has two main limitations. First the delay between the time when two nodes are physically able to exchange information and the time when they figure out that they are neighbors can be as high as the inter-beacon interval. This delay can be a major problem specially when mobiles are fast moving as the contact duration can be not enough for nodes to figure out that they are in contact. In order to reduce the impact of this effect, we need to reduce the beacon delay and send beacons frequently. However, in sparse networks most of the beacons are never received as connectivity is scarce. Increasing the rate of beacon transmission can result in the most of node battery to be consumed just for beacon transmission. 

The second issue is the dual problem. When the neighbor of a node moves and goes out of its reach it needs a time equal to disconnection timer delay to decide that the node is not anymore in the neighborhood. This means that a node might still continue to send message to a node that is not anymore in its neighborhood. This last point results in a large number of unsuccessful transmissions, that consumes node energy without having any use. The solution to this issue seems to reduce the disconnection delay, but as the disconnection delay should be larger to the beacon transmission interval this means reducing it and having the same issue as described before. 

This discussion shows that the beacon delay is an essential parameter controlling the performance of neighbor management and impacting strongly on the efficiency of the Epidemic Forwarding scheme.    
\subsection{Bloom filter management}\label{S:bfmanagement}
We explained before that each packet in the network is identified uniquely by a 4 bytes Packet ID. Now let us assume that a node has in its buffer $L$ packets and it wants to inform another encountered node about its buffer content. As described before several solutions for this problem have been proposed in the literature, among which Bloom filter \cite{Bloom}. Bloom filter is a data structure representing in a very compact way set membership. Therefore the Bloom filter can be used to represent the content of the buffer in a node. For this purpose, the node generates a Bloom filter that contains $L$ ($L\le N$) packet IDs and sends the information needed to retrieve this Bloom filter to the encountered node that use them to check the membership of its local packets and to decide which new packet to send back. The information needed to retrieve the Bloom filter consists in two components: first component is for retrieving the $K$ hashing function that are needed for the operation of the Bloom filter, and second component is a membership vector that contains the bits set by the hashing functions. For the first component we will assume that the node ID is used as the seed of a random generator that is used to generate random hash functions. Meaning that knowing the node ID enables to retrieve the hash functions used by this node. The membership vector is send directly to the neighbors. 

Bloom filters performance is controlled by the false alarm probability, {\em i.e.} the probability that an entry that is not in the set defined by the Bloom Filter is falsely reported as being in the set. This probability can be derived as a function of the number of hash functions $K$ and the length of the membership vector $M$ is a well known relationship. Generally one chooses $K$ in order to minimize the probability of false alarms. With this choice, if one wishes to insert $N$ values in the Bloom filter and set a target false alarm probability $p$, the minimal length of the membership vector $M$ and the needed number of hash functions that can achieve this false alarm rate is given as :
\begin{equation} \label{E:BF length}
	\begin{cases}
		M = -\frac {N \ln p }{ (\ln 2)^2}   
		\\
		K = \frac{M}{N}\ln 2\\
	\end{cases}	
\end{equation}
For example, if one wants to insert 50 packetIDs in its buffer in a Bloom filter and achieve a false alarm probability less than 2\% (one packet per Bloom filter), he will need $M=407,11$ bits and $K=5,6$ hash functions. By rounding these values, and aligning the memory to byte units, one needs 51 bytes to transfer the Bloom filter to neighbors in place of the 200 bytes needed to transfer 50 packet ID of 4 bytes each. The above Eq. shows that the memory requirements of Bloom filter increases with the number of inserted values. To control the size of the Bloom filter one has to limit the number of inserted values or to increase the acceptable false alarm rate. 

Nonetheless, over the time, the number of packets in the buffer of a node increases and therefore the number of values that have to be inserted in the Bloom filters increases, resulting in an linearly (with $N$) increasing Bloom filter size  or an increasing false alarm probability if the size is held constant. We have therefore to use a sliding window in order to manage the Bloom filter contents. A major tradeoff results between sliding window length (and the Bloom filter size) and size of probability of retransmitting a redundant packet (both resulting from a Bloom filter false alarm or from an active packet being dropped out of the window). This tradeoff represents the buffer management challenge we described in section II-A and has major performance impact on epidemic forwarding. In the forthcoming we will describe different strategies to deal with the above trade-off.
The first and trivial strategy, we will call strategy A, consists of choosing a maximal number of inserted values $N$ ($N$ can be larger than $L$ the node buffer size to account for packet that are destined to the node and have been removed from the buffer) and an acceptable false error probability $p$, and simply sending in each packet a Bloom filter containing the pktIDs of last $N$ packets received by the node. The receiving node uses these Bloom filters to detect which packets have been received by the neighbor in order to not forward them. Moreover, by checking the destination node ID of a packet with the reception status of the neighbor indicated by its Bloom filter, one can detect if a packet has reached its destination and to remove it from its buffer. In this case the trade-off is between the size $N$ (or equivalently the size of the Bloom filter) and the probability of a redundant packet. 

The second strategy, named B, consists of extending the idea of differential encoding to Bloom filter. Differential encoding has been widely used in compression and consists of first encoding completely an item and to send thereafter only the difference of the fully encoded item with its followers. This achieves a larger compression ratio as the differences can be encoded with much less bits than a full encoding. One can periodically fully re-encode an item in order to resynchronize. Extension of this idea to Bloom filters is straight forward. We assume that each beacon send periodically by the node to announce its presence, contains a big Bloom Filter containing $N$ entries, while each packet send by the node piggybacks a small Bloom filter containing $n<<N$ entries that have been added from the time of generation of the previous beacon. A node checking the reception of a packet by a neighbor will first check the small Bloom filter and check thereafter the big one if the first check is negative. This strategy achieves a better trade off than strategy A, but at the cost of an eventual desynchronization that results from beacons being send in larger interval times. 

The third strategy, named C, extend strategy B by observing that the data that have to be send in the Bloom filters consists in fact of two components: the buffer content that has a maximal size limit $L$, and the list of packets received by its destination that can grow indefinitely. We therefore extend strategy B by sending in beacons two Bloom filters: one containing $N \le L$ entries and one containing a list of $J$ last packets received by its destination. We will explain later how to construct in a distributed way this second list. The list of packets received by their destinations plays an important role as it acts as a collective acknowledge propagating in the network and freeing buffer space occupied by packets that have been delivered.

\subsection{Forwarding scheme}
We assume that the node maintains for each ``active" neighbor a data structure containing the set of Bloom filters provided by the neighbor (the small, large and received Bloom filter of strategy B and C above), a list of PktID named \emph{notReceivedYet} list containing the IDs of packets in the node buffers but not yet received at this neighbor, and information about the last packet forwarded to this neighbor. When a node has an opportunity (given by the scheduler or resulting from a contact) to send a packet to a neighbor, it first checks (using the last Bloom filters received from the neighbor) among the packet received from the time the last packet was forwarded to the neighbor which one are not received at the neighbor and adds them to the  \emph{notReceivedYet} list. Therefore, nodes can simply avoid forwarding duplicate packets to their neighbors by sending packets from their {\em notReceivedYet} lists. Among packets in this list packets that have as destination the neighbor have priority and will be forwarded first. Otherwise a randomly selected packet from the {\em notReceivedYet} is forwarded to the neighbor. A forwarded packet is removed from the \emph{notReceivedYet} list. To ensure synchronization and to manage packets not received by the destination, the  {\em notReceivedYet} is fully reconstructed  (by checking if any packet in the buffer is not received by the neighbor) whenever a beacon is received.

The scheduling between neighbors is also simply done by first checking if anyone of them is the final destination of a packet in the buffer. If it is the case the node is given priority. If no final destination exists among neighbors, one node is chose at random and packets are forwarded to it. The priority given to final destination node decreases the average delivery delay. However, provided no destination exists in the neighborhood, we choose one of the neighbors randomly and send a randomly selected packet to it to avoid discrimination. The sequence number of the sent packet should be removed from notReceivedYet of the receiver.

\section{Performance Evaluation} \label{S:evaluation}
In order to evaluate the performance of the proposed practical Bloom filter based epidemic forwarding methods, we implemented in the ns-3 simulation environment the proposed schemes. We are indeed aware of the shortage of simulation relative to representativeness and generalization, however up to our knowledge the Bloom filter based forwarding has not been implemented and tested, and we believe that the simulation environment can show weaknesses but needs indeed to be comforted with real deployment. In order to evaluate the proposed scheme we have used random waypoint mobility as well as the mobility resulting from San Francisco taxi traces \cite{SFtraces}. We are aware of the shortcoming of simulation based on random waypoint mobility described in \cite{le2005perfect} and we will therefore not draw our conclusions only based on these simulations but are evaluating also on more realistic traces like the San Francisco Taxi traces. 
\subsection{Random Waypoint mobility evaluation}
\begin{figure}[!ht]
\includegraphics[width=\columnwidth]{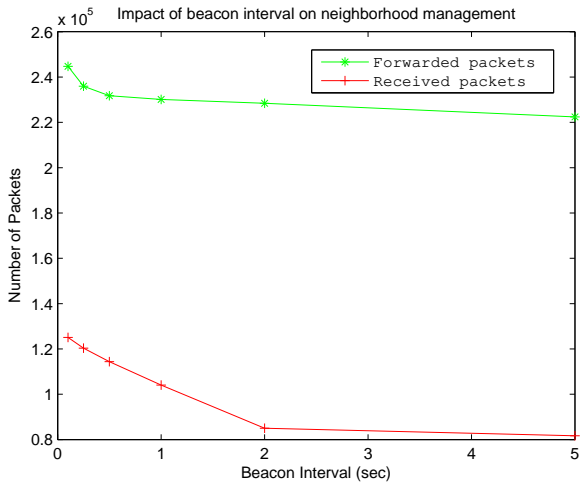}	
\caption{Impact of beacon delay on number of forwarded and received packets}
\end{figure}
Let us first evaluate the impact of neighbor management. We simulate 40 nodes moving in a $1000 \times 1000$ grid with a wifi transmission range of 50 unit each, {\em i.e.} around 30\% of the mobility area is covered by connectivity, with a rate of 1 Mbps. The node implements the neighborhood strategy we described before. We set the disconnection delay as 10\% larger than the beacon interval. We show in Fig. 1 the number of forwarded and received packets for different beacon delays ranging from 0.1 sec to 5 sec. As shown in Fig. 1, both the number of forwarded and received packets decreases while beacon delay increases. The gap between the two curves is the number of packets forwarded but that are not received by any node. The transmission efficiency is obtained by dividing the number of received packets by the number of forwarded packets. On this account, as Fig. 1 shows the efficiency decreases while the beacon delay increases. We observe the highest efficiency (51\%) when beacon delay is 0,1 seconds and the lowest efficiency (36\%) when beacon delay is assumed 5 seconds. From the curve we can see that for beacon delay larger than 2 seconds the efficiency drops severely. Based on the above curve we decided to set in the forthcoming a beacon delay of 1 sec. 
\begin{figure}
\includegraphics[width=\columnwidth]{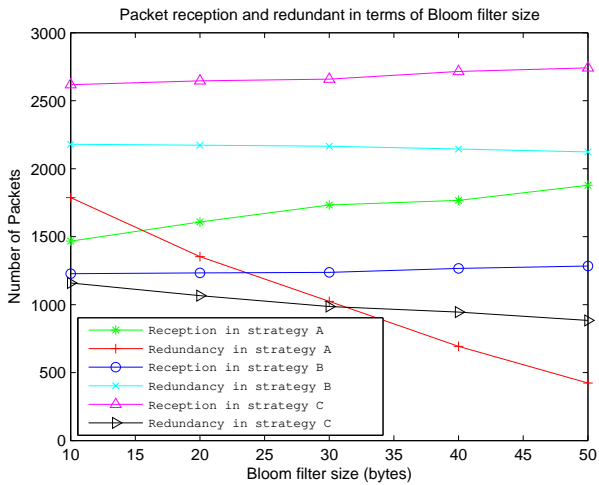}	
\caption{Comparison of number of received and redundant packets with different Bloom filter sizes for several Bloom filter management strategies}
\end{figure}
We thereafter evaluated on the same simulation scenario (40 nodes moving in a $1000 \times 1000$ grid with a wifi transmission range of 50 unit each), a scenario where each node acts simultaneously as source, relay and destination. The sources are assumed to be greedy meaning that whenever there is an opportunity to insert a new packet in the network they have data to send and insert it. We moreover assume that each packet has 1000 bytes of payload. The destination of messages is chosen randomly among all 40 nodes in the network. The buffer size of nodes is assumed to be 50 packets. Simulations are run 3600 seconds each. In strategy B, the beacons send Bloom filters that contain up to 200 packets and in strategy C the received packet Bloom filter has size 150 packets and the 50 packets in buffer are announced through a second Bloom Filter. All Bloom filters are designed with an objective of false alarm probability of 2\%. We compare in Fig. 2 the performance of different strategies in terms of number of packets received at final destinations and number of redundant packets while Bloom filter size increases. As shown in Fig. 2, by increasing the size of Bloom filter, performance improves in terms of both number of packets received at final destinations and number of redundant packets. Therefore, the trade off is between the Bloom filter size and its imposed overhead on information packets. We compare in Fig. 3 the performance in term of number of packets received at final destinations as a function of time. The three curves are relative to Bloom filter windowing strategy A, B and C. Interestingly the performance achieved by strategy A is better than strategy B. A more detailed analysis shows that strategy B achieves a much lower overhead (around 0.64\%) compared to strategy A (5\% of overhead). This shows itself by a larger number of redundant transmissions, {\em i.e.} 2.0\%, for strategy B, compared to 0.4\% for strategy A. However the delivery ratio, {\em i.e.} the number of delivered messages among generated ones for strategy B raise to 84\% while it was 71\% for strategy A. Nonetheless strategy C achieves a much better performance even if the redundancy is still larger than strategy A with 0.73 \%. This can be explained by the fact that in strategy C, we add a mechanism that propagates in the network the reception of packets by their final destinations and free the space occupied by them. Therefore the impact of congestion on this last strategy is less than for strategy A and B that have their buffer filled early in their lifetime. 
\begin{figure}
\includegraphics[width=\columnwidth]{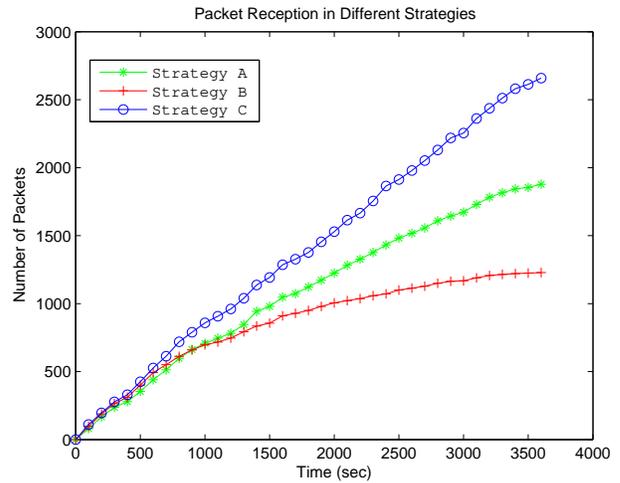}	
\caption{Comparison of number of packets received at destinations per unit of time for several Bloom filter management strategies for Random Waypoint mobility model}
\end{figure}
\subsection{San Francisco Trace evaluation}
\begin{figure}
\includegraphics[width=\columnwidth]{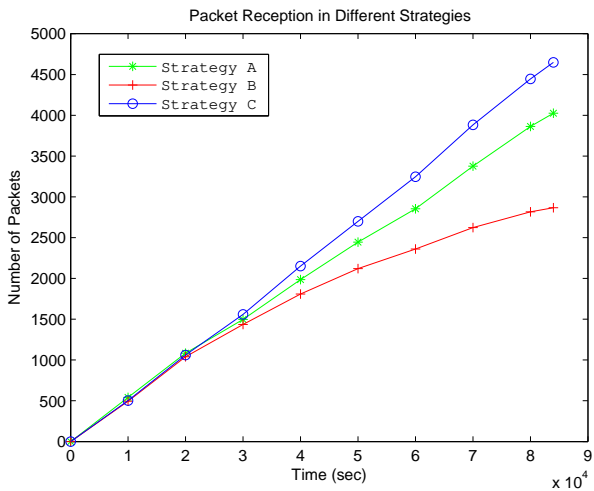}	
\caption{Comparison of number of packets received at destinations per unit of time for several Bloom filter management strategies for Taxi trace}
\end{figure}
We also run the same simulation as the one described above but with mobility data coming from San Francisco taxi GPS traces. We assumed here that the range of WIFI is 100 meters and that we have 100 taxis moving in the city. This scenario is sparser than the previous one and we had to run it for a longer time namely one day or equivalently 84 000 seconds. In order to be able to run this in a reasonable time we reduce the number of packets sent per second by a factor of 100 to reduce the number of events to simulate. We show in Fig. 4 the results in term of number of packets received at their final destinations. However as explained in order to run the simulation in tractable time we reduced the number of packets by a factor of 100. Short scale experiment we did without this reduction showed that one can expect to observe in reality around 53 times packet than predicted by the slowed simulation. However we present here the values as obtained directly from simulation without applying the 53 coefficient. The results obtained in Fig. 4 are very similar to what observed in Fig. 3. Interestingly we achieve a delivery ratio 75\% and that is very high accounting for the sparseness of the network. The achieved mean delay is around 200 minutes  that is relatively large but understandable with regards to the sparsity of the network. 

\section{Related Work} \label{S:rw}
Researchers have addressed some usages of Bloom filters in ad hoc networks. In \cite{bsub}, Zhao et al. present B-SUB (Bloom- filter-based pub-SUB system) as a content-based publish-subscribe system for the networks formed by human-carried wireless devices. The system employs a Temporal Counting Bloom Filter (TCBF) to perform content-based networking tasks. B-Sub mechanism also exploits TCBF in order to encode usersÕ interests and embed routing information. This results in a space efficient propagation of interests. Additionally, the authors analyze several methods for controlling the TCBFÕs false positive rate. In \cite{eiko}, Yoneki et al. propose another novel approach for content-based publish/subscribe system in mobile ad-hoc networks. Using aggregated summaries of content-based subscriptions in Bloom filters expression for the dynamic construction of an event dissemination structure, the authors extended ODMRP (On- Demand Multicast Routing Protocol). In \cite{hubaux}, Aad et al. propose novel packet coding techniques that make the combination possible, thus integrating the advantages in a more complete and robust solution. Those techniques exploit Bloom filters because they offer high compression rates, low false positives and no false negatives. In their case, the compression rate will help in reducing header sizes considerably, while (low) false positives have no considerable impact since they slightly increase the anonymity set size. The authors also analyzed the impact of false positives in their paper. In \cite{henderson}, Parris et al. investigate hiding social network information using one-way hashing, via the probabilistic Bloom filter data structure. In addition, The authors embedded the friends list within a Bloom filter instead of transmitting the senderÕs friends list as a list of nodes, in their Obfuscated Social Network Routing (OSNR) scheme.

\section{Conclusions} \label {S:conclusion}
We showed in this paper several strategies for implementing Bloom Filter based exchange of buffer content in epidemic forwarding schemes.   We evaluated the proposed schemes using an ns-3 simulation that validated the interest of the proposed techniques. We also addressed the issue of neighborhood management in DTNs which is a fundamental issue in these challenged networks and presented a neighborhood management scheme based on message exchange. We also showed that the efficiency of our neighborhood management scheme is dependent on two parameters namely beacon delay and disconnection delay. 


%
%
%



\bibliographystyle{IEEEtran}
%

%
%

\bibliography{myRefs}

\end{document}